\documentstyle[preprint,aps,graphicx,epsfig]{revtex}
\begin{document}
\def\unit{\hbox to 3.3pt{\hskip1.3pt \vrule height 7pt width .4pt \hskip.7pt
\vrule height 7.85pt width .4pt \kern-2.4pt
\hrulefill \kern-3pt
\raise 4pt\hbox{\char'40}}}
\def\II{{\unit}}
\def\cM {{\cal{M}}}
\def\half{{\textstyle {1 \over 2}}}

\def    \beq    {\begin{equation}} \def \eeq    {\end{equation}}
\def    \bea    {\begin{eqnarray}} \def \eea    {\end{eqnarray}}
\newcommand\mx[4]{\left#1\begin{array}{#2}#3\end{array}\right#4}

\preprint{UG-00-20}

\title{$Spin(p+1,p+1)$ Covariant D$p$--brane Bound States}
\author{P. Sundell\footnote{p.sundell@phys.rug.nl} }
\address{Institute for Theoretical Physics, University of Groningen,
\\ Nijenborgh 4, 9747 AG Groningen, The Netherlands }

\maketitle

\maketitle

\begin{abstract}

We construct $Spin(p+1,p+1)$ covariant D$p$--brane bound states by
using that the potentials in the RR sector of toroidically
compactified type II supergravity transform as a chiral spinor of
the T--duality group. As an application, we show the invariance of
the zero--force condition for a probe D--brane under noncommutative
deformations of the background, which gives a holographic proof of
the stability of the corresponding field theory ground state under
noncommutative deformations. We also identify the $Spin(p+1,p+1)$
transformation laws by examining the covariance of the D--brane
Lagrangians.

\end{abstract}

\section{Introduction}

Noncommutative extensions of geometry arise naturally in string
theory and M--theory at the boundaries of open
branes\cite{OpenBranes}. Such a boundary is restricted to end inside
some other 'host brane' where it experiences the force of a gauge
invariant modified field strength which is essentially given by the
pull--back to the host brane of a background potential. So even a
potential which has zero background field strength, may have
nontrivial effects on the open brane physics in the host brane. The
effect of the coupling of the boundary to the modified field
strength can be interpreted as that the open brane sees the host
brane as manifold with a noncommutative structure
\cite{NCYMlim,CH,us1}.

The dynamics of open brane--host brane systems tend to decouple from
the 'gravitational' dynamics in the bulk in critical limits, where
the electric components of the modified field strength and the
corresponding tensions diverge, leading to open branes with finite
effective tensions trapped inside the host branes. The idea to
decouple the gravitational dynamics from the dynamics on the brane
is an old one, as for instance \cite{MM,LST}. The essential new
ingredient is that by tuning the noncommutative deformation
parameters differently one can obtain different critical limits
giving rise to light degrees of freedom of various rank
\cite{NCOM,OBLST}. For example, a critical NS two--form leads to
decoupled open strings in a spatio--temporally noncommutative
D--brane \cite{NCOS}. Another example is a critical RR $(p-1)$--form
which leads to a decoupled open D$(p-2)$--brane in a D$p$--brane
\cite{NCOM,OBLST}. These systems have dual, perturbative
formulations in terms of Yang--Mills field theories on spatially
noncommutative D$p$--branes \cite{NCYMlim,CH,NCinst,NCYM,SW}, which
is the formulation of noncommutative geometry that first came to be
widely appreciated in the context of string theory. In M--theory a
critical three--form potential yields a decoupled open membrane
theory \cite{NCOM,us2} on a five--brane with noncommutative loop
space \cite{us1}. There are also noncommutative extensions of the NS
five brane little string theories \cite{NCOM,OBLST}.

The critical decoupling limits can be realized as ultra violet
limits for 'probe' host branes \cite{NCYMholo,OMholo,OBLST,BCOT} in
the near horizon regions of bound state solutions \cite{solutions}
consisting of a large numbers of 'source' host branes bound to
smeared sources for the critical fields. The advantage of this
holographic approach \cite{holo}, as opposed to simply considering
scaling limits in flat spacetime, is that the supergravity duals
incorporate essential nonperturbative aspects of the decoupled
theories. Thus, a spacetime duality transformation that relates two
bound state solutions induces a transformation between the two
corresponding supergravity duals and their related ultra violet
limits, that in turn leads to a duality between the decoupled
theories \cite{NCOM,OBLST,us2}. Therefore, the various
noncommutative theories form a duality web which is inherited from
the underlying spacetime theory. As a result, all maximally
supersymmetric noncommutative theories in less than or equal to six
dimensions can be traced back to the noncommutative M--theory
five--brane \cite{NCOM,OBLST}.

This far, however, most aspects of the noncommutative open brane
theories have been found by examining the (effective) probe brane
field theories \cite{NCinst,NCDBI,NCYMft,probing}. From this point
of view, the spacetime background provides the field theory moduli.
A noncommutative deformation of the field theory corresponds to
turning on a given (smeared) source in the bound state. The bound
state incorporates the nonperturbative back--reaction on the
remaining sectors, such as the tensions and the dilaton which are
dual to the field theory moduli space metric and coupling.
Therefore, the holographic setup provides a useful nonperturbative
tool for examining the effect of noncommutative deformations on the
field theory moduli \cite{CH,probing,TdualNCDbrane,MultiTh}. In this
paper we shall exploit this in showing the deformation independence
of the zero--force condition for a probe brane; see also
\cite{newpaper}.

To facilitate this, we shall first obtain a uniform expression for
all deformed D$p$--brane bound state solutions related to a given
undeformed D$p$--brane solution\footnote{We note that there are
several more or less equivalent solutions in the literature starting
with \cite{solutions}, but as we shall see the particular form that
we obtain here is more well--suited for the calculations of the
brane potential.}. We recall that the noncommutative deformations of
a type II D$p$--brane unify with the 'diagonal' T--duality
transformations into the T--duality group $O(p+1,p+1)$. Actually,
having non--zero flux in a $p$--brane bound state breaks the
$ISO(p,1)$ covariance. It is therefore natural to wrap the brane on
a $(p+1)$--dimensional torus, which may be uncompactified after the
deformation. Note, the decoupling limits are actually sensitive to
whether the brane world volume is compact or not \cite{Danielsson},
but this distinction will not be important at the level of the
effective field theory that we shall examine in this paper. Since
the full string theory is $O(p+1,p+1;{\bf Z})$ invariant
\cite{TDnonpert}, while the low energy effective actions possess
$O(p+1,p+1;{\bf R})$ symmetry \cite{Maharana,Fukuma}, it follows
that the space of inequivalent noncommutative moduli is given by
$O(p+1,p+1;{\bf R})/O(p+1,p+1;{\bf Z})$ (indeed the discrete
symmetry acts as a certain equivalence, known as Morita equivalence,
between noncommutative field theories on D--branes \cite{Morita}).
Thus, the desired bound states can be obtained by acting with the
corresponding $O(p+1,p+1)$ group element on the undeformed solution.

The massless sector of toroidically compactified type II string
theory, which transforms irreducibly under the U--duality group,
splits under the T--duality subgroup $O(d,d)$ into tensor
representations in the NS sector and spinor representations in the
RR sector \cite{SenVafa}. The transformations in the RR sector were
first found in \cite{RRTD}, and were later related to the
'covariance' of the D$p$--brane Lagrangians \cite{DbrTD} under
T--duality transformations. It was then realized \cite{Dpot} that
the form of the transformations in the RR sector simplifies using a
new set of RR potentials. Later, guided by this and \cite{SenVafa},
the authors of \cite{Fukuma} gave a manifestly $Spin(d,d)$ invariant
form of the RR sector of the toroidically compactified supergravity
Lagrangian (including the CS terms). The work of \cite{Hassan}
included fermions and massive IIA deformations by examining the
'twist' between the left and right--moving local Lorentz frames on
the string worldsheet induced by the T--duality. A similar approach
was taken by the authors of \cite{CLPS}. Recently the
$Spin(p+1,p+1)$ covariant formulation of D--brane actions with
extended anomaly polynomials was given by \cite{HassanMinasian}.

In Section 2 we examine the above mentioned $Spin(d,d)$ covariant
formulation of D--branes. To some extent we review old results, but
we also give a useful alternative derivation of the $Spin(d,d)$
transformations in the RR sector by requiring covariance of D--brane
actions, in the spirit of the work of \cite{DbrTD}.

In Section 3 we give a manifestly $Spin(p+1,p+1)$ covariant
parameterisation of D$p$--brane bound states, and use this
construction to examine some effects of noncommutative deformations
on the field theory moduli space. It is found that the brane
potential is only renormalized by a multiplicative factor, which
gives a holographic proof of the stability of the field theory
ground state under noncommutative deformations.

We end with a discussion in Section 4.

\section{$Spin(d,d)$ Covariance of D--branes}

In this section we discuss the T--duality transformations of the
bosonic field content in toroidically compactified type II
supergravity. Two natural ways of finding them are by examining
either the invariance of the full supergravity Lagrangian
\cite{Maharana,Fukuma,Fukuma}, or the 'covariance' of brane sigma
models \cite{DbrTD,Hassan,CLPS,Buscher,brTD}. By covariance we here
mean that the sigma model actions in two T--dual background are
equal up to some duality transformation of the world volume fields.
The covariance of the NS string world sheet action, which involves
dualising a scalar in two dimensions, was used in this context to
derive the T--duality transformations in the NS sector
\cite{Buscher}. These were then incorporated into a manifestly
$O(d,d)$ invariant low energy effective supergravity Lagrangian
\cite{Maharana}. The covariance of D--branes, which involves double
dimensional reductions, was used to derive the T--duality
transformations in both NS and RR sectors \cite{DbrTD}. Later these
were incorporated into a manifestly $Spin(d,d)$ invariant
supergravity Lagrangian by the work of \cite{Fukuma,Hassan,CLPS}. In
particular, this requires a new set of RR potentials \cite{Dpot}
that transform linearly under $Spin(d,d)$. Below we shall reexamine
the derivation of \cite{DbrTD}, using the new potentials and some
other convenient techniques of \cite{Fukuma} and
\cite{HassanMinasian} for writing the Wess--Zumino term, and verify
that the covariance of D--branes indeed yields the expected
$Spin(d,d)$ transformations the RR sector.

\subsection{General setup}

Let us begin by setting the notation and recalling the covariance properties
of D--branes under target space T--duality transformations. Thus we consider
the action of a D$p$--brane with tension $T_p$:

\beq S_p[X,A;E,\phi;D]=T_p\left(S_{\rm DBI}[X,A;E,\phi]+S_{\rm
WZ}[X,A;D]\right)\ ,\label{action}\eeq \beq S_{\rm
DBI}[X,A;E,\phi]=-\int d^{p+1}\xi
e^{-\phi}\sqrt{-\det\left((f^*E)_{\alpha\beta}+\alpha'F_{\alpha\beta}\right)}\
.\label{dbi}\eeq \beq S_{\rm WZ}[X,A:D]=\int e^{\alpha'F}\wedge
f^*D\ ,\label{wz}\eeq

Here the fields on the brane are the embedding
coordinates\footnote{We denote the ten--dimensional spacetime
indices by $M,N,\dots=0,\dots,9$; indices of $T^d$ by
$\mu,\nu\dots=0,\dots,d-1$; the uncompactified spacetime indices by
$i,j,\dots=d,\dots ,9$; and the world volume indices of a
D$p$--brane by $\alpha,\beta,\dots=0,\dots,p$. Importantly, $T^d$
may contain the time--direction as for example in the case of a
brane configuration.} $X^M(\xi^\alpha)$, with pull--back denoted by
$f^*$, and the world volume vector field $A_{\alpha}$ with field
strength $F=dA$. The background fields in the NS sector are the
dilaton $e^\phi$ and the metric and two--form potential which have
been combined in

\beq E_{MN}=g_{MN}+B_{MN}\ .\eeq

In the RR sector the potential $D$ for the RR field strength $R$ is
defined by \cite{Fukuma,Dpot}:

\beq dR=-dB\wedge R\ ,\quad R=e^{-B}\wedge dD\ .\label{dpot}\eeq

We use multi--form notation; for example

\beq D=\sum_K\frac1{K!}D_{M_1\cdots M_1}dx^{M_1}\wedge\cdots\wedge
dx^{M_K}\ .\eeq

The Bianchi identity can also be solved using a $C$ potential
related to $D$ as follows:

\beq R=dC+dB\wedge C\ ,\quad C=e^{-B}\wedge D\ .\label{cpot}\eeq

Note, the $D$ potential is not invariant under NS gauge
transformations. However, the transformation reads $\delta D= \delta
B\wedge D$, which is in fact only to expected if $D$ transforms
linearly under $O(d,d)$, since constant NS gauge transformations on
$T^d$ is a subset of the generators of $O(d,d)$.

To find the full set of $O(d,d)$ transformations $(E_{MN},
\phi;D)\rightarrow (\tilde{E}_{MN},\tilde{\phi};\tilde{D})$ we
examine the $O(d,d)$ covariance of the D--branes \cite{DbrTD}, i.e.

\beq
S_{\tilde{p}}[\tilde{X},\tilde{A};\tilde{E},\tilde{\phi};\tilde{D}]=
S_p[X,A;E,\phi;D]\ ,\label{cov} \eeq

under the following three basic types of $O(d,d)$ transformations
\cite{Morita} \footnote{We coordinitize $O(d,d)$ by $2d\times 2d$
matrices \beq \Lambda=\mx{(}{ll}{a&b\\c&d}{)}\ ,\quad
\Lambda^TJ\Lambda=J\ ,\eeq where $J$ is the special $O(d,d)$ element
given by $a=d=0$, $b=c={\bf 1}$. This presentation is equivalent to
orthogonal matrices $\tilde{\Lambda}=U\Lambda U^{-1}$;
$U=\frac1{\sqrt2}(1+i\sigma^2)\otimes {\bf 1}$, obeying
$\tilde{\Lambda}^T\eta\tilde{\Lambda}=\eta$,
$\eta=\sigma^3\otimes{\bf 1}$.}:

\begin{enumerate}

\item[i)] The $GL(d)$ generators

\beq \Lambda_R=\mx{(}{ll}{R^{-1}&0\\0&R^T}{)}\ ,\quad R\in GL(d)\
,\label{glR} \eeq

with the following action on the brane data:

\beq \tilde{p}=p\ ,\quad \tilde{T}_{\tilde{p}}=T_p\ , \eeq \beq
\tilde{X}^i=X^i\ ,\quad \tilde{X}^\mu=X^\nu  R_\nu{}^\mu\ ,\quad
\tilde{A}_\alpha=A_\alpha\ . \label{lambdar} \eeq

\item[ii)] The generators of constant NS gauge transformations

\beq \Lambda_{b}=\mx{(}{ll}{1&b\\0&1}{)}\ ,\quad
b_{\mu\nu}=-b_{\nu\mu}\ , \label{cns}\eeq

with the following action on the brane data:

\beq \tilde{p}=p\ ,\quad \tilde{T}_{\tilde{p}}=T_p\ , \eeq \beq
\tilde{X}^M=X^M\ ,\quad
\tilde{F}_{\alpha\beta}=F_{\alpha\beta}-(\alpha')^{-1}(f^*b)_{\alpha\beta}\
. \label{lambdath} \eeq

\item[iii)] The 'diagonilised T--duality generator'

\beq
\Lambda_{\hat{\mu}}=\mx{(}{ll}{1-e_{\hat{\mu}}&-e_{\hat{\mu}}\\-e_{\hat{\mu}}&
1-e_{\hat{\mu}}}{)}\ ,\quad
(e_{\hat{\mu}})_{\nu\rho}=\delta_{\hat{\mu}\nu}\delta_{\hat{\mu}\rho}\
,\quad \hat{\mu}=0,\dots,d-1\ , \label{lambdam} \eeq

whose action on the brane data is given by a double dimensional
reduction followed by an 'uplift' as follows:

\beq \tilde{p}=p-1\ ,\quad \tilde{T}_{\tilde{p}}=R_{\hat{\mu}} T_p\
, \eeq \beq \partial_p X^M=\mx{\{}{ll}{0&\mbox{for }M\neq
\hat{\mu}\\ 1&\mbox{for }M=\hat{\mu}}{.}\ ,\quad \partial_p
A_\alpha=0\ , \label{ddr3} \eeq\beq
\tilde{X}^M=\mx{\{}{ll}{X^M&\mbox{for }M\neq\hat{\mu}\\
-\alpha'A_p&\mbox{for }M=\hat{\mu}}{.}\ ,\quad
\tilde{A}_{\tilde{\alpha}} = A_\alpha\ ,\quad
\tilde{\alpha}=0,\dots,p-1\ .\label{ddr5} \eeq

where $R_{\hat{\mu}}$ is the radius of the $\hat{\mu}$'th
direction.

\end{enumerate}

\subsection{Transformations in the NS Sector}

We first briefly review the results for the NS sector; for a fuller
review see e.g. \cite{TDnonpert}. The NS sector is KK decomposed as
follows\cite{Maharana,Fukuma}

\bea
E_{\mu\nu}&=&g_{\mu\nu}+B_{\mu\nu}\ ,\label{kksc}\\
E_{\mu i}&=&E_{\mu\nu}A^\nu_i-B'_{i,\mu}\ ,\\
E_{i\mu}&=&A_i^\nu E_{\nu\mu}+B'_{i,\mu}\ ,\\
E_{ij}&=&E'_{ij}+E_{\mu\nu}A^\mu_i A^\nu_j+B'_{[i|,\mu}A_{j]}^\mu\
, \label{kktens} \eea

where $E_{\mu\nu}$ are the KK scalars; the KK vectors are given by

\beq A^r_i=(B'_{i,\mu},A^\mu_i)\ ,\quad
r=({}_\mu,{}^\mu)=1,\dots,2d\ ;\label{kkvect}\eeq

and $E'_{ij}=g'_{ij}+B'_{ij}$ are the KK tensors. From (i) and (ii)
it follows that under the generators $\Lambda_R$ and $\Lambda_b$,
the KK scalars (\ref{kksc}) transform as

\beq \tilde{E}=R^{-1}E R^{-T}\ ,\quad \tilde{E}=E+b\ ,\label{fl1}
\eeq

where we have used matrix notation to suppress the $\mu,\nu$
indices, and the KK vector (\ref{kkvect}) transform as

\beq \tilde{A}_i=\Lambda_R A_i\ ,\quad \tilde{A}_i=\Lambda_b A_i\
,\label{kkv1} \eeq

where we have suppressed the $r,s$ indices. It then follows that the
KK tensor $E'_{ij}$ is invariant under $R$ and $b$ transformations.
To analyze (iii) one may use the Jordan decomposition formula for
determinants. It follows that under the generator
$\Lambda_{\hat{\mu}}$ the KK scalars and vectors transform as

\beq
\tilde{E}=E+\frac1{g_{\hat{\mu}\hat{\mu}}}(1-E)e_{\hat{\mu}}(1+E)\
,\label{fl2} \eeq \beq \tilde{A}_i=\Lambda_{\hat{\mu}}\tilde{A}\
,\label{kkv2} \eeq

while the KK tensor is invariant. Hence, under a general $O(d,d)$
transformation with generator

\beq \Lambda=\mx{(}{ll}{a&b\\c&d}{)}\ , \label{odd}\eeq

it follows from (\ref{fl1}) and (\ref{fl2}) and the fact that an
arbitrary $O(d,d)$ element can be built by a series of
transformations of the form (i)--(iii), that the KK scalars undergo
the fractional linear transformation

\beq \tilde{E}=(aE+b)(cE+d)^{-1}\ .\label{flt} \eeq

This is equivalent to the bivector representation
$\tilde{M}=(\Lambda^{-1})^T M \Lambda^{-1}$, where $M$ is the
$O(d,d)$ element

\beq M_{rs}=\mx{(}{ll}{g^{\mu\nu}&-g^{\mu\rho}B_{\rho\nu}\\
B_{\mu\rho}g^{\rho\nu}& g_{\mu\nu}-B_{\mu\rho}g^{\rho\sigma}B_{\sigma\nu}}{)}
\ .\label{bivector} \eeq

>From (\ref{kkv1}) and (\ref{kkv2}) it follows that the KK vectors transform as
an $O(d,d)$ vector:

\beq \tilde{A}_i=\Lambda A_i\ , \eeq

while finally the KK tensor is inert:

\beq \tilde{E}_{ij}=E_{ij}\ . \eeq

The dilaton is inert under $\Lambda_R$ and $\Lambda_b$, while under
$\Lambda_{\hat{\mu}}$ it follows from the Jordan decomposition
formula that
$e^{-\tilde{\phi}}=e^{-\phi}\sqrt{|g_{\hat{\mu}\hat{\mu}}|}$. This
is equivalent to

\beq e^{\tilde{\phi}}=e^\phi\left({\frac{\det
\tilde{g}_{\mu\nu}}{\det g_{\mu\nu} }}\right)^{\frac14}\
,\label{dil} \eeq

where we have used (\ref{fl2}) to compute $\det\tilde{g}_{\mu\nu}=
(g_{\hat{\mu}\hat{\mu}})^{-2}\det g_{\mu\nu} $. Since the ratio of
determinants in the right side of (\ref{dil}) is inert under
$\Lambda_R$ and $\Lambda_b$ it follows that (\ref{dil}) is actually
true for general $O(d,d)$ transformations, which concludes the
analysis of the NS sector.

\subsection{Transformations in the RR Sector}

The analysis of the RR sector becomes more transparent by using
the isomorphism between the algebra of exterior line elements
$dX^M$ and inner derivatives $i_M$ and the following auxiliary
fermionic oscillator algebra \cite{Fukuma}:

\beq \{\hat{\psi}_M,\hat{\psi}^{\dagger N}\}=\delta_M^N\ ,\quad
\{\hat{\psi}_M,\hat{\psi}_N\}=0=\{\hat{\psi}^{\dagger
M},\hat{\psi}^{\dagger N}\}\ ,\label{stosc} \eeq \beq
(\hat{\psi}_M)^{\dagger}=g_{MN}\hat{\psi}^{\dagger N}\ , \eeq

acting on the vacuum $|0\rangle$, which is annihilated by
$\hat{\psi}_M$, and its hermitian conjugate
$\langle0|=(|0\rangle)^{\dagger}$. Mapping a $K$--form $\Omega_K$
to the operator

\beq \widehat{\Omega}_K=\frac {(-1)^{\frac{K(K-1)}{2}}} {K!}
\hat{\psi}^{\dagger M_1}i_{M_1}\cdots \hat{\psi}^{\dagger
M_K}i_{M_K}\Omega_K\ , \label{hatO}\eeq

one has the isomorphism $\Omega\wedge \Xi\mapsto
\widehat{\Omega}\widehat{\Xi}$. The WZ term (\ref{wz}) can now be
written as \cite{HassanMinasian}

\beq S_{\rm WZ}=\int \langle\omega_p|e^{\alpha'\widehat{F}}|
D\rangle\ ,\label{wzosc} \eeq

\beq \mbox{where}\qquad \langle \omega_p|=\langle 0|
f^*\left(dX^{M_1}\wedge\cdots \wedge dX^{M_{p+1}}\right)
\hat{\psi}_{M_{p+1}}\cdots \hat{\psi}_{M_1}\ ,\eeq \beq \widehat{F}=\frac12
F_{\alpha\beta} \hat{\psi}^{\dagger \alpha}\hat{\psi}^{\dagger \beta}\ ,\quad
|D\rangle=\widehat{D}|0\rangle\label{hatf} \eeq

and we have defined the pull--back to the brane of the space--time
oscillator algebra (\ref{stosc}) as follows:

\beq \hat{\psi}_\alpha=\partial_\alpha X^M\hat{\psi}_M\ ,\quad
\{\hat{\psi}_\alpha,\hat{\psi}^{\dagger\beta}\}=\delta_\alpha^\beta\
,\quad \{\hat{\psi}^{\dagger
\alpha},\hat{\psi}^{\dagger\beta}\}=0\ . \label{wvosc}\eeq

The last two conditions are of course equivalent to
$\hat{\psi}_\alpha=(f^*g)_{\alpha\beta}(\hat{\psi}^{\dagger\beta})^{\dagger}$.
The volume form $\langle\omega_p|$ acts as an
'intertwiner' between the two oscillator algebras (\ref{stosc}) and
(\ref{wvosc}):

\beq \langle
\omega_p|\left(\hat{\Omega}-\widehat{f^*\Omega}\right)=0\
.\label{itw}\eeq

Under the $GL$ transformation (i) we have

\beq d\tilde{X}^M\psi_M=\widehat{\Lambda}_R ~dX^M\hat{\psi}_M
~\widehat{\Lambda}_R^{-1}\ , \quad
\hat{\tilde{\psi}}_\alpha=\partial_\alpha\tilde{X}^M\hat{\psi}_M=
\widehat{\Lambda}_R \hat{\psi}_\alpha \widehat{\Lambda}_R^{-1}\ ,
\eeq

where the oscillator representation of $R_\mu{}^\nu\equiv
(\exp(A))_\mu{}^\nu$ is given by

\beq \widehat{\Lambda}_R=\sqrt{\det R}\exp (-\hat{\psi}^{\dagger
\mu}A_\mu{}^\nu \hat{\psi}_\nu)\ . \eeq

Using $\langle0|\widehat{\Lambda}_R=\langle0|$, it follows that

\beq
\langle\tilde{\omega}_p|=\langle\omega_p|\widehat{\Lambda}_R^{-1}\
,\quad
\widehat{\tilde{F}}=\widehat{\Lambda}_R\widehat{F}\widehat{\Lambda}_R^{-1}\
,\eeq

such that the requirement of covariance, eq. (\ref{cov}), implies that the
generator $\Lambda_R$ is represented on the RR potential as

\beq |\tilde{D}\rangle =\widehat{\Lambda}_R|D\rangle\ .\label{lr}
\eeq

Under the NS gauge transformation (ii) the volume form
$\langle\omega_p|$ is inert, while

\beq \langle\omega_p| e^{\alpha'\widehat{\tilde{F}}} =
\langle\omega_p| \exp(-\frac12
(f^*b)_{\alpha\beta}\hat{\psi}^{\dagger \alpha} \hat{\psi}^{\dagger
\beta}) e^{\alpha'\widehat{F}} = \langle\omega_p| \exp(-\frac12
b_{\mu\nu}\hat{\psi}^{\dagger \mu} \hat{\psi}^{\dagger
\nu})e^{\alpha'\widehat{F}}\ ,\eeq

where we have used (\ref{itw}). Hence it follows from (\ref{cov})
that the generator $\Lambda_\theta$ is represented on the RR
potentials as

\beq |\tilde{D}\rangle =\widehat{\Lambda}_b|D\rangle\ ,\quad
\widehat{\Lambda}_b = \exp(\frac12 b_{\mu\nu}\hat{\psi}^{\dagger
\mu} \hat{\psi}^{\dagger \nu})\ .\label{lth} \eeq

Finally, for the diagonilised T--duality (iii) we compute

\bea&& \langle\omega_p|e^{\alpha'\widehat{F}}\nonumber\\&=&
d\xi^{p}\wedge\langle0|\left[f^*(dX^{\check{M}_1} \wedge\cdots
\wedge dX^{\check{M}_{p}})\hat{\psi}_{\check{M}_p}\cdots
\hat{\psi}_{\check{M}_1}\right]\hat{\psi}_{\hat{\mu}}~
e^{\alpha'\partial_{\tilde{\alpha}} A_{p}\hat{\psi}^{\dagger
\tilde{\alpha}}\hat{\psi}^{\dagger
p}}e^{\alpha'\widehat{\tilde{F}}}\nonumber\\
&=&d\xi^{p}\wedge\langle0|\left[f^*(dX^{\check{M}_1} \wedge\cdots
\wedge dX^{\check{M}_{p}})\hat{\psi}_{\check{M}_p}\cdots
\hat{\psi}_{\check{M}_1}\right]\widehat{\Lambda}_{\hat{\mu}}~
e^{\alpha'\partial_{\tilde{\alpha}} A_{p}\hat{\psi}^{\dagger
\tilde{\alpha}}\hat{\psi}^{\dagger
p}}e^{\alpha'\widehat{\tilde{F}}}\nonumber\\
&=&d\xi^{p}\wedge\langle0|\left[f^*(dX^{\check{M}_1} \wedge\cdots
\wedge dX^{\check{M}_{p}})\hat{\psi}_{\check{M}_p}\cdots
\hat{\psi}_{\check{M}_1}\right]e^{\partial_{\tilde{\alpha}}
\tilde{X}^{\hat{\mu}}\hat{\psi}^{\dagger
\tilde{\alpha}}\hat{\psi}_{\hat{\mu}}}
 ~e^{\alpha'\widehat{\tilde{F}}}~\widehat{\Lambda}_{\hat{\mu}}\nonumber\\
&=&d\xi^{p}\wedge\langle0|\tilde{f}^*(dX^{M_1} \wedge\cdots \wedge
dX^{M_{p}})\hat{\psi}_{M_p}\cdots \hat{\psi}_{M_1}
~e^{\alpha'\widehat{\tilde{F}}}~\widehat{\Lambda}_{\hat{\mu}}\nonumber\\
&=& d\xi^{p}\wedge
\langle\tilde{\omega}_p|~\widehat{\Lambda}_{\hat{\mu}}\ ,
\label{comp}\eea

where $\check{M}$ is an auxiliary nine--dimensional index which
skips $\hat{\mu}$, and we have defined

\beq
\hat{\Lambda}_{\hat{\mu}}=\hat{\psi}_{\hat{\mu}}+\hat{\psi}^{\dagger
\hat{\mu}}\ , \quad \hat{\Lambda}_{\hat{\mu}}^2=1\ . \eeq

We have also used
$\widehat{\Lambda}_{\hat{\mu}}\hat{\psi}^{\tilde{\alpha}\dagger}\hat{\psi}^{\dagger p}
\widehat{\Lambda}_{\hat{\mu}}=-\hat{\psi}^{\tilde{\alpha}\dagger}\hat{\psi}_{\hat{\mu}}$ and
the chain rule, which yields:

\beq e^{-\partial_{\tilde{\alpha}}
\tilde{X}^{\hat{\mu}}\hat{\psi}^{\dagger
\tilde{\alpha}}\hat{\psi}_{\hat{\mu}}}
f^*(dX^{\check{M}})\hat{\psi}_{\check{M}}
e^{\partial_{\tilde{\alpha}}
\tilde{X}^{\hat{\mu}}\hat{\psi}^{\dagger
\tilde{\alpha}}\hat{\psi}_{\hat{\mu}}}
=\tilde{f}^*(dX^M)\hat{\psi}_M\ . \eeq

Here $f$ denotes the embedding of the original doubly dimensional
reduced D$(p-1)$--brane resulting from (\ref{ddr3}) and $\tilde{f}$
denotes the embedding of the uplifted $(p-1)$--brane that results
from identifying the T--dual direction with the internal vector
field as in (\ref{ddr5}). Since $\int d\xi^p=R_{\hat{\mu}}$, it
follows from (\ref{cov}) that the generator $\Lambda_{\hat{\mu}}$ is
represented on the RR potentials as

\beq |\tilde{D}\rangle=\hat{\Lambda}_{\hat{\mu}}|D\rangle\
.\label{lm} \eeq

One can show \cite{Fukuma} that Eqs. (\ref{lr}), (\ref{lth}) and
(\ref{lm}) defines a reducible $Spin(d,d)$ representation, see also
\cite{Hassan}, such that under a general $O(d,d)$ transformation
with generator $\Lambda$ given by (\ref{odd}) we have

\beq |\tilde{D}\rangle = \widehat{\Lambda}|D\rangle\ .
\label{Dtr}\eeq

In terms of the $C$ potential, eq. (\ref{Dtr}) amounts to the
transformation law \cite{DbrTD,RRTD}

\beq |\tilde{C}\rangle = e^{-\widehat{\tilde{B}}}~\widehat{\Lambda}
e^{\widehat{B}}~|C\rangle\ .\label{ctr}\eeq

The explicit expression for the $Spin(d,d)$ generator
$\widehat{\Lambda}$ is given in \cite{Fukuma}, but it suffices for
our purposes here to note that

\beq \widehat{(\Lambda^T)}=\widehat{\Lambda}^{\dagger}\
.\label{dagger}\eeq

We remark that the $Spin(d,d)$ representation decomposes into
$2^{d-1}$--dimensional chiral irreps of fixed form degree on the
transverse space, and with total form degree being either odd for
IIA, which yields negative chirality, or even for IIB, which yields
positive chirality . For example, when $d=2$ ($\mu=0,1$) the KK
decomposition of the even IIB potentials leads to scalars
$\{D_{(0)}, D_{01}\}$; one--forms $\{D_{(1)i\mu}\}$; two--forms
$\{D_{(2)ij}, D_{(4)ij01}\}$ and so on. The RR sector of the
supergravity Lagrangian, including the CS term, can be written
\cite{Fukuma} in terms of the $D$ potential as a sum of kinetic
terms $R\wedge\star_{g_{MN}}R$ that has a manifestly $Spin(d,d)$
invariant KK decomposition in terms of spinor bilinears
$\bar{R}^A\wedge \star R^B S_{AB}(M)$ which employs the spinor
element $S_{AB}(M)$ formed out of the $O(d,d)$ bivector element
(\ref{bivector}).

\section{Application to Noncommutative Theories}

In this section we first use the $O(p+1,p+1)$ and $Spin(p+1,p+1)$
transformations of the NS and RR sectors discussed in the previous
section to write a $Spin(p+1,p+1)$ covariant form of the D$p$--brane
solutions to type II supergravity (we are going to take the IIA mass
deformation to be zero which effectively restricts our analysis to
$p<7$ \cite{massiveIIA}). We then examine these solutions using a
brane probe, and show that the probe brane potential is
multiplicatively renormalized under a noncommutative deformation. We
remark that other similar aspects of noncommutative deformations of
field theories have been studied recently in the holographic setup;
see e.g. \cite{TdualNCDbrane}.

\subsection{$Spin(p+1,p+1)$ covariant bound state}

We start from an $ISO(p,1)$ invariant D$p$--brane configuration

\beq ds^2=g_{\mu\nu}dx^\mu dx^\nu + g_{ij} dx^i dx^j\ ,\quad
B_2=\beta_2\ ,\quad e^{\phi}\ ,\label{bc1}\eeq \beq C=\omega
dx^0\wedge\cdots dx^p\wedge (1+\alpha)+\gamma\label{bc2} \eeq

where $\omega$ is a function and $\beta_2$, $\gamma$ and $\alpha$
are transverse forms:

\beq i_\mu\beta_2=i_\mu\alpha=i_\mu\gamma=0\ . \label{imu}\eeq

Here $i_\mu$ is the inner derivative in the $\mu$'th direction. We
compactify the world volume in all $d=p+1$ directions and use a
$Spin(d,d)$ transformation to generate a new solution and then
decompactify. In case some directions are kept constant, the
deformed solution gives rise to a supergravity dual of a
noncommutative wound string theory \cite{Danielsson}.

The transformation leading to a supergravity dual of a theory with
noncommutativity parameter $\theta^{\mu\nu}$ of dimension
(length)$^2$ is generated by the group element

\beq \Lambda=\Lambda_{\hat{0}}\cdots \Lambda_{\hat{p}}
\Lambda_{-\frac{\theta}{\alpha'}} \Lambda_{\hat{p}}\cdots
\Lambda_{\hat{0}}=\mx{(}{ll}{1&0\\
\frac{\theta}{\alpha'}&1}{)}=\left(\Lambda_{\frac{\theta}{\alpha'}}\right)^T\
,\label{ge}\eeq

with the spinor representation given by (\ref{lth}) and (\ref{dagger})

\beq \widehat{\Lambda}=\widehat{\Lambda}_{\hat{0}}\cdots
\widehat{\Lambda}_{\hat{p}}
\widehat{\Lambda}_{\frac{\theta}{\alpha'}}
\widehat{\Lambda}_{\hat{p}}\cdots
\widehat{\Lambda}_{\hat{0}}=\left(\widehat{\Lambda}_{\frac{\theta}{\alpha'}}\right)^{\dagger}=
\exp\left(\frac{\theta^{\mu\nu}}{2\alpha'}\hat{\psi}_\mu
\hat{\psi}_\nu \right)\ .\label{sr}\eeq

This corresponds to first T--dualising in all brane directions, such
that one ends up with an array of smeared D$(-1)$--branes, then
switching on constant NS fluxes on the T--dual torus, and finally
T--dualising back to a D$p$--brane again. Note that an arbitrary
$O(d,d)$ group element $\Lambda$ can be decomposed into a rotation
$\Lambda_R$, a constant gauge transformation $\Lambda_b$ and a
constant T--dual gauge transformation $(\Lambda_{\tilde{b}})^T$,
such that $\Lambda=(\Lambda_{\tilde{b}})^T\Lambda_R\Lambda_b$.
However, a shift of the NS two--form potential in two directions
that are tangent to a D--brane has no physical effect on the open
string theory on the D--brane, so that $\Lambda_b$ is a trivial
deformation of the open string theory on the D$p$--brane. Nor has
$\Lambda_R$ any physical effect. Hence, since we wish to deform the
open string theory on a probe brane, there is no loss of generality
to assume (\ref{ge}).

The deformed configuration, which we denote by tildes, is determined
by using (\ref{flt})--(\ref{dil}) in the NS sector (there are no KK
vectors), and (\ref{ctr}) in the RR sector:

\bea\tilde{g}_{\mu\nu}&=&g_{\mu\rho}\left[(1-(\alpha')^{-2}(\theta
g)^2)^{-1}\right]^\rho{}_\nu\ ,\qquad\qquad \tilde{g}_{ij}=g_{ij}\
,\label{dcs}\\
\tilde{B}_{\mu\nu}&=&g_{\mu\rho}\theta^{\rho\sigma}g_{\sigma\lambda}
\left[(1-(\alpha')^{-2}(\theta g)^2)^{-1}\right]^\lambda{}_\nu\
,\quad \tilde{B}_{ij}=\beta_{ij}\ ,\\
e^{\tilde{\phi}}&=&{e^{\phi}\over \sqrt{\det(1-(\alpha')^{-2}(\theta
g)^2)}}\ ,\eea \beq
\tilde{C}=e^{-\frac12\tilde{B}_{\mu\nu}dx^\mu\wedge dx^\nu}\left(
\omega e^{\frac1{2\alpha'}\theta^{\mu\nu}i_\mu i_\nu}
dx^0\wedge\cdots\wedge dx^o(1+\alpha)+\gamma\right)\ .
\label{tildec}\eeq

The above solution is equivalent to many other solutions in the
literature \cite{solutions,NCYMholo,MultiTh}. From the expression
for $\tilde{B}$ and $\tilde{C}$, we see that by choosing $\theta$ in
two spatial directions, say $1$ and $2$, we create a D$p$--D$(p-2)$
bound state, where the $(p-2)$--brane is in the directions
$3,\dots,p$ \cite{NCYMholo}. By instead choosing a spatio--temporal
$\theta$ in, say, the $0$ and $1$ directions, we create a D$p$--F$1$
bound state with the string in the $1$ direction
\cite{NCYMholo,NCOS}. By using many blocks in $\theta$ we can create
multi--brane bound states \cite{MultiTh}.

We distinguish between the following two different situations: 1)
magnetic deformations, for which $-(\theta g)^2$ is positive; 2)
electric deformations, for which $-(\theta g)^2$ has two negative
eigenvalues. Magnetic deformations yield non--singular
configurations, while electrically deformed configurations are
non--singular provided that the electric field is not too large. For
example, in the case that the undeformed configuration is an
extremal brane which interpolates between a near horizon region and
an asymptotic region, such that we can take
$g_{\mu\nu}=H^{-\frac12}\eta_{\mu\nu}$, with $0<H^{-\frac12}<1$,
then an electrically deformed configuration is non--singular
provided the electric eigenvalue $-\theta_{\rm e}$ obeys
$\theta_{\rm e}\leq \alpha'$.

\subsection{Supergravity Duals}

As mentioned in the introduction, the near horizon regions of
D$p$--brane bound states are supergravity duals of open string
theories living on probe D$p$--branes \cite{holo,NCYMholo}. In the
ultra violet region of the near horizon region, where the separation
between the probe and the stack of source branes (given in rescaled
horizon units) goes to infinity, the open string dynamics at the
probe decouples from the bulk closed string dynamics and defines an
ultra violet complete theory on $p+1$ dimensions without gravity.

In the case of an electrically deformed solution, the supergravity
dual is obtained by the limit $\theta_{\rm e}\rightarrow \alpha'$
keeping everything else fixed in units of $\alpha'$. The resulting
'electric' near horizon region thus includes both the near horizon
region and the asymptotic region of the original undeformed brane
configuration. The asymptotic geometry is however that of a smeared
string, which leads to a critical decoupling limit in this region.
The fact that the critical decoupling limit takes place in the
asymptotic region in a way explains why noncommutative string theory
does not require the presence of D--branes, as was recently pointed
out by \cite{Danielsson}.

For a magnetically deformed D$p$--brane the near horizon limit is
instead identical to the one of the original undeformed
configuration\footnote{ This is the field theory near horizon limit
\cite{holo} where $x^\mu$, $\Phi^i\equiv x^i/\alpha'$,
$ds^2/\alpha'$, $B/\alpha'$ and $C_p/(\alpha')^{\frac{p}2}$ are
fixed. Here $\Phi^i$ are the scalar fields with canonical dimension
one. In the UV limit of the near horizon region,
$g_{\mu\nu}/\alpha'$ and $\omega/(\alpha')^{\frac{p+1}2}$ diverges,
while all transverse forms and line elements goes to zero, since the
gradients $\partial_\mu \Phi^i$ are fixed.} with the additional
condition that the (magnetic) deformation parameter
$\theta^{\mu\nu}$ is kept fixed. Hence, field theory--near horizon
limits and magnetic deformations commute, while this is not true in
the electric case. This is equivalent to the fact that field
theories do not admit spatio--temporal noncommutativity deformations
\cite{Rabi1}. Further aspects of this are discussed in
\cite{newpaper}.

The magnetically deformed D$p$--brane supergravity duals thus give
rise to decoupled noncommutative field theories (which are ultra
violet complete in four dimensions; in higher dimensions the
completion is a decoupled open string or brane theory and the field
theory is an effective description for slowly varying fields). The
field theory excitations are confined to the vanishing locus of the
probe brane potential $V$. The reason is that excitations that break
the zero--force condition are 'frozen out' in the decoupling limit,
as their potential energy scales to infinity in the limit. The
picture is that of a potential valley that gets increasingly narrow
at higher energy scales, with a curvature at the bottom that
diverges in the ultra violet. Therefore the zero--force condition is
one of the basic moduli of the noncommutative theory, and it is
natural to ask whether this data will depend on the noncommutativity
parameter?

\subsection{A Holographic Non--Renormalization Theorem for the Field
Theory Ground State}

In order to examine the effects on the zero--force condition of a
noncommutative deformation, we expand the Lagrangian of a probe
D$p$--brane in the magnetic near horizon region of the deformed
background (\ref{dcs}-\ref{tildec}) in the UV. Going to the static
gauge and using (\ref{dil}) and

\beq
\tilde{g}_{\mu\nu}+\tilde{B}_{\mu\nu}=\left[(g^{-1}+\frac{\theta}{\alpha'})^{-1}
\right]_{\mu\nu}\ ,\quad
\det\left(\tilde{g}_{\mu\nu}+\tilde{B}_{\mu\nu}\right) =\sqrt{\det
\tilde{g}_{\mu\nu} \det g_{\mu\nu}}\ ,\eeq

we find that the contribution from the kinetic term (\ref{dbi}) to
the potential has the expansion\footnote{ The tension
$T_p=(\alpha')^{-\frac{p+1}2}\mu_p$. In the near horizon region the
powers of $\alpha'$ are absorbed into fixed tensions and fluxes, and
the dimensionless charge density $\mu_p$ is fixed. The field
strength $F_{\mu\nu}$ and $\mu_p$ are fixed in the UV.}:

\bea &&T_p ~
e^{-\tilde{\phi}}\sqrt{-\det(\tilde{g}_{\mu\nu}+\tilde{B}_{\mu\nu}+\alpha'
F_{\mu\nu})}
\nonumber\\ &=&T_pe^{-\phi}\sqrt[4]{{\det g_{\mu\nu}\over \det
\tilde{g}_{\mu\nu}}} \sqrt{-\det
(\tilde{g}_{\mu\nu}+\tilde{B}_{\mu\nu})}
\sqrt{\det(1+(\alpha'g^{-1}+\theta)F)}\nonumber\\ &=&T_p
e^{-\phi}\sqrt{-\det g_{\mu\nu}} \sqrt{\det(1+\theta F)}+{\rm
finite} .\label{fin}\eea

The finite bit is an expansion in traces of the form

\beq T_p e^{-\phi}\sqrt{-\det
g_{\mu\nu}}(\alpha')^2tr\left[g^{-1}Fg^{-1}F(\theta F)^n\right]\ ,
\label{traces}\eeq

that build up the kinetic term for the noncommutative gauge field
and scalars \cite{SW,newpaper}. Using the oscillator formalism in
(\ref{hatO}) and (\ref{wzosc}), and the transformation law
(\ref{Dtr}) together with (\ref{sr}), we find that the contribution
from the Wess--Zumino term (\ref{wz}) to the potential is given by:

\bea &- T_p&\!\!\!\!\langle 0| d\xi^0\wedge\cdots \wedge
d\xi^p~\hat{\psi}_p\cdots\hat{\psi}_0 ~e^{\frac12
F_{\mu\nu}\hat{\psi}^{\dagger\mu}\hat{\psi}^{\dagger\nu}}
e^{\frac12\frac{\theta^{\mu\nu}}{\alpha'}\hat{\psi}_\mu\hat{\psi}_\nu}
~\omega \hat{\psi}^{\dagger 0}\cdots\hat{\psi}^{\dagger
p}|0\rangle
\nonumber\\
&=&-T_p~d^{p+1}\xi ~\omega ~\langle 0|
e^{\frac12\frac{\theta^{\mu\nu}}{\alpha'}\hat{\psi}_\mu\hat{\psi}_\nu}
e^{\frac12
F_{\mu\nu}\hat{\psi}^{\dagger\mu}\hat{\psi}^{\dagger\nu}}
|0\rangle \nonumber\\
 &=& -T_p~ d^{p+1}\xi ~\omega~\sqrt{\det(1+\theta
F)}\ .\label{vev} \eea

Note, the determinant is a square of Pfaffians, so that the last
expression is indeed a polynomial in traces of $\theta F$. A
noncommutative deformation therefore leads to a multiplicative
renormalization of the potential, i.e. the deformed potential

\beq \tilde{V}=\sqrt{\det(1+\theta F)}~V\ ,\label{tilv} \eeq

\beq \mbox{where}\qquad V=T_p\left(e^{-\phi}\sqrt{-\det{g}}- \omega\right)
\label{pot}\eeq

is the undeformed potential. We remark that the dilaton scales in
the UV precisely as to make (\ref{traces}) finite, and hence the
first term in (\ref{fin}) diverges (note that $\theta$ and $F$ are
fixed). Moreover, $T_p\omega$ diverges in the UV, so (\ref{vev})
diverges in the UV too. Hence, if non--zero, the potential
(\ref{pot}) diverges, so that in the UV the excitations of the field
theory are confined to the vanishing locus of (\ref{pot}) which
defines the ground state of the field theory. Hence, from
(\ref{tilv}) it follows that the noncommutative deformation does not
shift the field theory ground state.

\section{Discussion}

The non--renormalization theorem for the zero--force condition has
consequences for the S--duality between noncommutative Yang--Mills
and open string theories (for $p=3$), which will be reported
elsewhere \cite{newpaper}. Here it suffice to mention that in a case
with less symmetry, such that $V=0$ is a non--trivial truncation of
the probe brane field theory, as is e.g. the case with brane
solutions which approach warped AdS compactifications in their near
horizon regions \cite{warped}, it follows from (\ref{tilv}) that the
exact Gross--Nekrasov monopole solution \cite{GN} can be embedded in
the UV--complete field theory on the vanishing locus.

The deformed solution (\ref{dcs})--(\ref{tildec}) reproduce many
solutions already known in the literature. In particular, since the
$O(d,d)$ transformation does not break supersymmetry, it follows
that by starting from a maximally symmetric D--brane solutions and
considering general $\theta$ that one can have bound states with
several different smeared sources that still has $16$ unbroken
supersymmetries. For instance, the interesting 'maximal'
$D4-D2-D2-D0$/$M5-M2-M2-MW$ IIA/M--theory solution \cite{BCOT}, that
appeared only recently and that has a very interesting phase
structure, can be obtained rather straightforwardly from
(\ref{dcs})--(\ref{tildec}) by starting with a maximally symmetric
D$4$--brane and taking non--zero $\theta^{12}$ and $\theta^{34}$.
The duality chain of maximal M--theory/IIB solutions will give rise
to an enlarged duality web of decoupled noncommutative theories as
well as ten and eleven--dimensional theories, and the deformed
solution (\ref{dcs})--(\ref{tildec}) may be useful tool in examining
this \cite{wip}

\section*{Acknowledgements}

I am grateful to discussions with Eric Bergshoeff, David Berman,
Martin Cederwall, Henric Larsson and Bengt Nilsson. Part of this
work will also appear in a joint publication \cite{newpaper}. My
research is supported in full by Stichting FOM.

\end{document}